\documentclass[aps,reprintnumbers,amsmath,amssymb,twocolumn]{revtex4}

\usepackage{graphicx}
\usepackage{dcolumn}
\usepackage{bm}
\usepackage{color}
\usepackage{amssymb}   

\usepackage{amsthm}

\newcommand{\rv}{{\mathbf r}}

\newcommand{\vel}{{\bf v}}

\newcommand{\keyw}[1]{\textbf{\textit{Keywords: }}#1}

\begin{document}

\author{A. Scacchi}
\author{J.M. Brader}
\email{joseph.brader@unifr.ch}
\affiliation{Department of Physics, University of Fribourg, CH-1700 Fribourg, Switzerland}

\title{Local phase transitions in driven colloidal suspensions}

\pacs{pacs numbers here.}

\begin{abstract}
Using dynamical density functional theory and Brownian dynamics simulations 
we investigate the influence of a driven tracer particle on the density distribution of a 
colloidal suspension at a thermodynamic statepoint close to the liquid side of the binodal. 
In bulk systems we find that a localized region of the colloid-poor phase, a `cavitation bubble', forms 
behind the moving tracer. The extent of the cavitation bubble is investigated as a function 
of both the size and velocity of the tracer.
The addition of a confining boundary enables us to investigate the interaction between the local 
phase instability at the substrate and that at the particle surface. 
When both the substrate and tracer interact repulsively with the colloids we observe the 
formation of a colloid-poor bridge between the substrate and the tracer. 
When a shear flow is applied parallel to the substrate the bridge becomes 
distorted and, at sufficiently high shear-rates, disconnects from the substrate to form 
a cavitation bubble.  
\end{abstract}

\maketitle

\keyw{Colloidal suspension, local phase transition, cavitation, shear flow.}


\section{Introduction}
Local information about the nonequilibrium properties of complex fluids can be obtained 
from microrheology experiments \cite{mason}. 
In passive microrheology, the thermal motion of a mesoscopic probe particle is tracked, typically 
using microscopy or diffusing wave spectroscopy, to obtain its mean-squared displacement. 
From this, one can infer the viscoelastic moduli 
by means of a generalized Stokes-Einstein relation. 
In the case of active microrheology, the tracer displacement is measured as it is 
pulled through the medium with a given force, typically implemented using 
either laser tweezers or magnetic fields. 
The force-displacement relation thus obtained represents a local analogue of a standard 
constitutive equation relating the macroscopic stress to the strain 
\cite{macosko,larson,mewis}.  

Nonlinear effects in active microrheology occur when the forces imposed by the tracer 
on the surrounding colloids are sufficiently large that they change the local 
microstructure, i.e.~the spatial 
distribution of colloids in the vicinity of the driven tracer becomes significantly 
different from that in equilibrium. 
Strongly nonlinear phenomena can occur when the system is in a state for which 
the motion of the tracer can cause a local phase change. 
Under such sensitive conditions, 
small changes in the tracer force can lead to large microstructural changes. 
An example of such behaviour is presented by the delocalization transition in 
dense, glassy suspensions; when sufficient force is applied to an embedded tracer a 
local devitrification occurs, leading to a sudden 
enhancement of the tracer mobility \cite{gazuz,gnann};

An alternative possibility, which will be the focus of the present work, is when the suspension 
is at a bulk thermodynamic statepoint close to, but outside, the binodal. 
In previous work we have demonstrated that the motion of the tracer can induce a local liquid-vapour 
phase transition, leading to the formation of either a `bubble' of the colloid-poor phase in a 
dense system, or a `trail' of colloid-rich phase in a low density system \cite{micro}.  
The former phenomenon is the colloidal analogue of the well-known cavitation effect in 
molecular liquids, e.g.~the bubbles formed by a propellor rotating in water.  

The dynamical density functional theory (DDFT), derived from the Langevin equation \cite{marconi_tarazona_1,marconi_tarazona_2}, or from the Smoluchowski equation \citep{archer_evans}, provides a very convenient method to 
study the influence of the equilibrium phase boundary on the dynamical 
response of the system (see also \citep{archer_rauscher}). 
The adiabatic approximation 
underlying DDFT enables direct connection to be made between the dynamics and an underlying 
equilibrium free energy functional. This makes possible an unambiguous identification of 
the position, with respect to the phase boundary, of the thermodynamic statepoint of interest. 
In Ref.~\cite{micro} we used DDFT to investigate colloidal cavitation 
with a minimal two-dimensional model. We refer to Refs. \cite{hopkins, malijevski, malijevski_parry} for related works. 

A repulsive interaction between the tracer and the (mutually attractive) colloidal 
particles leads to a drying layer around the 
tracer as the binodal is approached from the liquid side, although the geometrical constraint of the curved tracer surface renders this layer 
finite. Under such conditions, pulling the tracer at a constant velocity was found to generate 
a cavitation bubble behind the tracer. 
Although the basic phenomenology of colloidal cavitation was established in Ref.~\cite{micro}, 
we investigated neither the dependence of the cavitation on the model parameters nor did we provide 
independent confirmation of the effect using many-body simulations. 
Moreover, the influence of the confining boundaries was neglected. 

In this paper we will address these issues to provide a more complete picture of 
the colloidal cavitation phenomenon. 
In addition to providing an exploration of the parameter space we will also 
consider the more realistic situation where the tracer particle lies close to a 
confining wall. 
Such a situation occurs close to the container walls in sedimentation experiments. 
We will investigate the influence of the boundaries on the density profile around a 
nearby tracer and study how the cavitation bubble becomes modified. 
The same two-dimensional model as used in \cite{micro} will be employed. 
The paper is structured as follows: 
in section \ref{theory} we set out the theoretical basis for our investigations: 
in \ref{MicroscopicDynamics} we outline the many-body Brownian dynamics and DDFT, 
in \ref{approximation} we introduce the approximate density functional required for 
a closed theory, 
in \ref{numerics} 
we provide details of our numerical algorithms and in \ref{model} we describe the specific 
colloidal model under consideration. 
In \ref{Two dimensional cavity} we present the numerically calculated two dimensional density 
distributions for several different tracer sizes. 
In \ref{Flat wall capillarity} we study the behaviour of the density around a tracer close 
to a flat hard wall. 
In section \ref{Brownian dynamics simulations} we provide a brief investigation of cavitation 
using Brownian Dynamics simulations. 
Finally, in section \ref{discussion} 
we discuss our findings and give suggestions for future work.


\section{Theory}\label{theory}
\subsection{Microscopic dynamics and dynamical density functional 
theory}\label{MicroscopicDynamics}
We consider $N$ interacting, Brownian particles, suspended in a Newtonian solvent. 
In the absence of hydrodynamic interactions the time evolution of the 
configurational probability distribution is given by the Smoluchowski 
equation \cite{dhont_book} 
\begin{equation}
\frac{\partial \Psi(\textbf{r},t)}{\partial t}+\sum_{i=1}^{N}\nabla_i\cdot \textbf{j}_i(\textbf{r},t)=0\label{smolu_1}
\end{equation}
where the current of particle $i$ is given by
\begin{equation}
\textbf{j}_i(\textbf{r},t)=\textbf{v}_i^s(\textbf{r},t)\Psi(\textbf{r},t)
-D_0(\nabla_i-\beta F_i(\textbf{r}))\Psi(\textbf{r},t)\label{smolu_2}
\end{equation}
where $\textbf{v}_i^s(\textbf{r},t)$ is the solvent velocity, $D_0$ is the bare diffusion 
coefficient, $\beta=(k_BT)^{-1}$ and the force on the particle $i$ is generated by the pair 
potential $\phi(r)$ according to 
$\textbf{F}_i(\textbf{r})=-\nabla_i\sum_{j\neq i}\phi(r_{ij})$. 
Integrating Eq.~(\ref{smolu_1}) over all but one particle coordinates generates 
an exact coarse-grained equation for the one-body density
\begin{equation}\label{eq:eom_rho1}
\frac{\partial \rho({\bf r},t)}{\partial t} = -\nabla\cdot {\bf J}(\rv,t), 
\end{equation}
where the one-body particle flux is given by 
\begin{align}\label{flux}
{\bf J}(\rv,t)&=D_0\rho(\rv,t)\Bigg(
\,\nabla \ln\rho({\bf r},t)  
+\beta\nabla V_{\rm ext}({\bf r},t) \notag\\
&\hspace*{-0.5cm}-D_0^{-1}\vel^{\rm s}(\rv,t)+\beta\int \!d{\rv}' \frac{\rho^{(2)}({\bf r},{\bf r}',t)}{\rho(\rv,t)}
\nabla {\phi (\tilde{r})}
\Bigg), 
\end{align}
with the equal-time, nonequilibrium two-body density, $\rho^{(2)}({\bf r},{\bf r}',t)$, 
and where $\tilde{r}=|\rv - \rv'|$.  
In equilibrium, the integral term in \eqref{flux} obeys exactly the sum rule \cite{evans79}
\begin{equation}\label{sumrule}
\int\!d\rv'\,  \frac{\rho(\rv,\rv')}{\rho(\rv)}\nabla {\phi(\tilde{r})}
= \nabla\frac{\delta \mathcal{F}^{\rm exc}[\rho({\bf r})]}{\delta \rho(\rv)}, 
\end{equation}
where $\mathcal{F}^{\rm exc}[\,\rho({\bf r})]$ is the excess part of the Helmholtz free energy 
functional, dealing with the interparticle interactions.				
The total Helmholtz free energy is composed of three terms:
\begin{align}
  \mathcal{F}[\rho(\textbf{r})]=\mathcal{F}^{\rm id}[\rho(\textbf{r})] 
+ \mathcal{F}^{\rm exc}[\rho(\textbf{r})] +
  \int \!d\rv\, \rho(\rv)V_{\rm ext}(\rv),
\label{helmholtz}  
\end{align}
where the ideal gas part is given exactly by 
\begin{equation}\label{ideal}
\mathcal{F}^{\rm id}[\,\rho(\rv)]=\int d{\bf r}\, \rho(\rv)[\,\ln(\Lambda^3\rho(\rv))-1\,], 
\end{equation}
where $\Lambda$ is the thermal wavelength.		
To obtain dynamical information about the system, it is neccessary to make an adiabatic approximation 
by assuming that equation (\ref{sumrule}) holds also in non-equilibrium. 
This approximation allows to write the standard DDFT equation of motion 
\begin{equation}\label{stddft}
\frac{\partial\rho(\textbf{r},t)}{\partial t}+\nabla\cdot(\rho(\textbf{r},t)\textbf{v}^{s}(\textbf{r},t))=D_0\nabla\left[\rho(\textbf{r},t)\nabla
\frac{\beta\delta\mathcal{F}[\rho]}{\delta\rho}\right].
\end{equation}
By making the adiabatic approximation one thus obtains a dynamical theory which is closed on the 
level of the one-body density. 
The appearance of the affine solvent flow on the left hand side of Eq.~\eqref{stddft} is a consequence 
of making the adiabatic approximation. In a more complete theory the particle interactions will 
also modify the flow field and generate non-affine motion \cite{kb1,kb2,skb}. 



\subsection{Van der Waals approximation}\label{approximation}
Assuming that $\rho^{(2)}(\textbf{r},\textbf{r}')=\rho(\textbf{r})\rho(\textbf{r}')$ we can 
split excess free energy into a sum of a reference term and a mean-field perturbation. 
The excess free energy is then given by
\begin{align}
  \mathcal{F}^{\rm exc}[\rho]=\mathcal{F}^{\rm exc}_{\rm hd}[\rho] 
+ \frac{1}{2}\int\! d\rv_1\!\! \int\! d\rv_2 \,\rho(\rv_1)\rho(\rv_2)\phi_{\rm att}(r_{12}). 
\label{meanfield}  
\end{align}
In this work we will focus on two-dimensional systems interacting via a pair potential consisting of 
a hard-core repulsion and an attractive tail, $\phi_{\rm att}(r_{12})$, where $r_{12}=|\rv_1-\rv_2|$. 
We take the attraction to be zero within the hard-core. 
The excess free energy of the reference system $\mathcal{F}^{\rm exc}_{\rm hd}[\rho]$ is taken to be 
that of hard-disks, for which highly accurate approximations exist \cite{oettel}. 
This ensures that the equilibrium density profile generated by our theory is in good 
agreement with simulation data. 
The mean field approximation \eqref{meanfield} is known to provide qualitatively good results 
for many equilibrium problems. The performance of the mean field functional has very recently been reassessed \cite{archer_chacko}.

\subsection{Numerics}\label{numerics}

The density profile of colloids around a circular tracer is not circularly symmetric when 
either an external flow or a potential boundary is present. This necessitates a full 
two-dimensional solution of Eq.~\eqref{stddft} for which a careful choice of numerical 
grid is essential to obtain satisfactory numerical accuracy. 
For this reason we use methods similar to the Finite Element Method (FEM),
which provides a great deal of flexibility; a very fine grid can be used in sensitive 
regions and a coarse grid where the density is slowly varing. 
This keeps the computational demand low and allows the treatment
of relatively large systems, which becomes necessary at larger tracer velocities 
in order to accomodate the extended wake structures which develop.

A main complication is the evaluation of nonlocal terms stemming from the excess free energy
contribution. This is a problem which does not occur in standard finite element-type approaches 
and requires a custom-made solution; Fast-Fourier-Transform techniques 
are not applicable \cite{roland_review}. 
Fortunately, the integration kernels required for both the hard-disk reference 
system and the mean field term in \eqref{meanfield} 
can be precalculated.  
The finite range of the weight functions enables memory requirements to be kept 
at manageable levels.
Our custom numerics we have developed are based on the finite element framework 
deal.II \cite{Bangerth2007} and the grids used for our calculations were created 
automatically using Gmsh \cite{Geuzaine2009}.


\subsection{Model system and parameters}\label{model}
The model we will investigate is a two-dimensional system interacting 
via an attractive square well pair potential with the form 

\begin{align}\label{pair_potential}
\phi(r) = \left\{
  \begin{array}{lr}
    \infty \hspace*{5mm}\;&  0 < r < \sigma\\
    -\epsilon   & \sigma \le r < \sigma\delta\;\,\\
    0  &    r \ge \sigma\delta
  \end{array}
\right.
\end{align} 

where $\epsilon$ is the depth of the attractive well, $\delta$ sets the range of the attraction and 
$\sigma\equiv 2R$ is the diameter of the hard-disks. 
Henceforth, all distances will be measured in units of the disk radius $R$.

As only relative velocities are of physical significance, the motion of a tracer pulled through 
a bulk suspension is implemented by setting the solvent velocity field equal to a constant value 
in the $x$-direction, 
$\vel^{\rm solv}=-v^{\rm solv}\hat{\bf e}_{\rm x}$. 
We will henceforth use a dimensionless tracer velocity, 
defined according to
\begin{align}\label{bare_peclet}
v \equiv\frac{v^{\rm solv}R}{D_0}, 
\end{align} 
which compares the timescale of external driving, $\tau_{\rm drive}\equiv R/v_{\rm solv}$, 
with that of diffusive relaxation, $\tau_{\rm diff}\equiv R^2/D_0$. 
For $v>1$ external driving will dominate diffusive relaxation and the system will be pushed 
out of equilibrium.   

When the tracer lies close to the wall surface the velocity 
of the suspension is taken to be
\begin{align}\label{gradient_flux}
\textbf{v}(x,y)=y\>v\>\hat{\bf e}_{\rm x}
\end{align} 
which has a zero-component in the $y$-direction. 
%
The square-well system is characterized by two parameters, $\epsilon$ and $\delta$. 
In the following we will fix the range of the potential to the value $\delta=2R$ and report 
our numerical results in terms of the attraction strength parameter
\begin{align}\label{B}
B = -4(\delta^2-1)\epsilon\varphi,
\end{align}
where the two-dimensional area fraction is given by $\varphi=\pi N/V$, where 
$N$ is the number of particles and $V$ is the system volume.   
The parameter $B$ emerges naturally from the bulk limit of the mean-field density functional 
\eqref{meanfield} and enables the phase behaviour of the system for all values of $\epsilon$ 
and $\delta$ to be captured by a single phase diagram in the $(\varphi\,,B)$ plane. 

\begin{figure}[t!]
\includegraphics[scale=0.11]{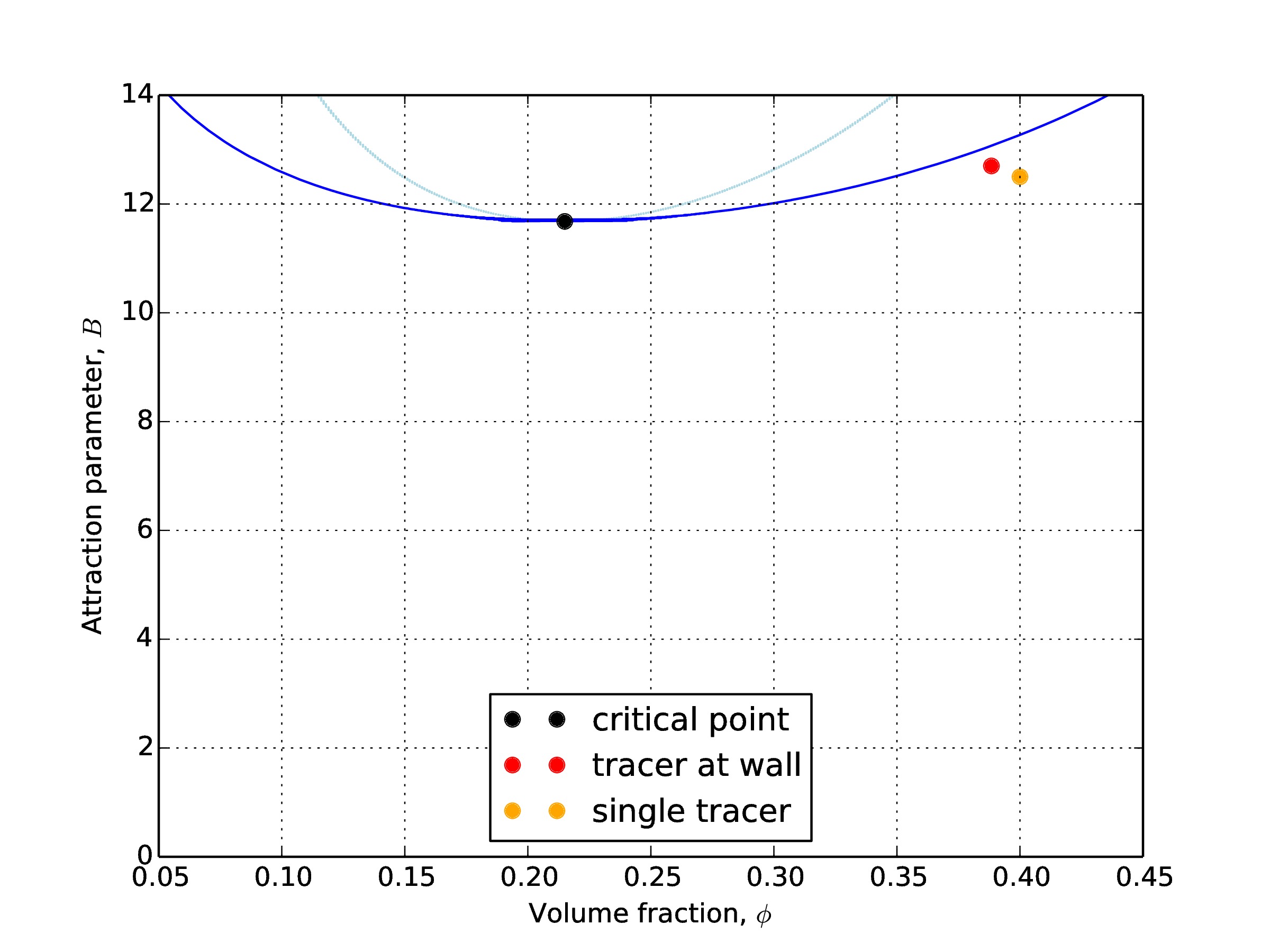}
\caption{In dark blue, the binodal line, in light-blue the spinodal line. For a ($\varphi$, B) coordinate system, we report in black the critical point (0.215, 11.68), in orange the one for the single tracer study (0.4, 12.5) treated in section \ref{Two dimensional cavity} and in red, the thermodynamic state point for the wall-tracer problem (0.3884, 12.7) in section \ref{Flat wall capillarity}.}\label{phase_diagram}
\end{figure}

For sufficiently high attraction strength ($B>B_{\rm crit}\approx 11.68$) the bulk free energy, obtained by setting $\rho(\rv)=\rho_{\rm b}$ in \eqref{ideal} and \eqref{meanfield}, exhibits a van der Waals loop, indicating the onset of gas-liquid phase separation. In Fig. \ref{phase_diagram} we show the phase diagram of the square well system, including both the binodal line enclosing the coexistence region and the spinodal line which marks the boundary of mechanical stability.  
\\

\section{Results}\label{results}

We consider here the response of the square-well host suspension to a purely repulsive tracer. 
The system, consisting of a tracer plus a colloidal suspension, is fully specified by the pair potential \eqref{pair_potential} together with the tracer-colloid interaction potential, given by:
\begin{align}\label{dry_potential}
u_{\rm tc}(r) = \left\{
  \begin{array}{lr}
    \infty &\;\; 0 < r < R_{c}\\
    0  &\;\;  r \ge R_{c}
  \end{array}
\right.
\end{align}
where we define $R_{c}=R_{t}+R$ the contact radius, which is the sum of the tracer radius $R_{t}$ and the radius of the colloidal disks $R$. 

\begin{figure}[t!]
\includegraphics[scale=0.18]{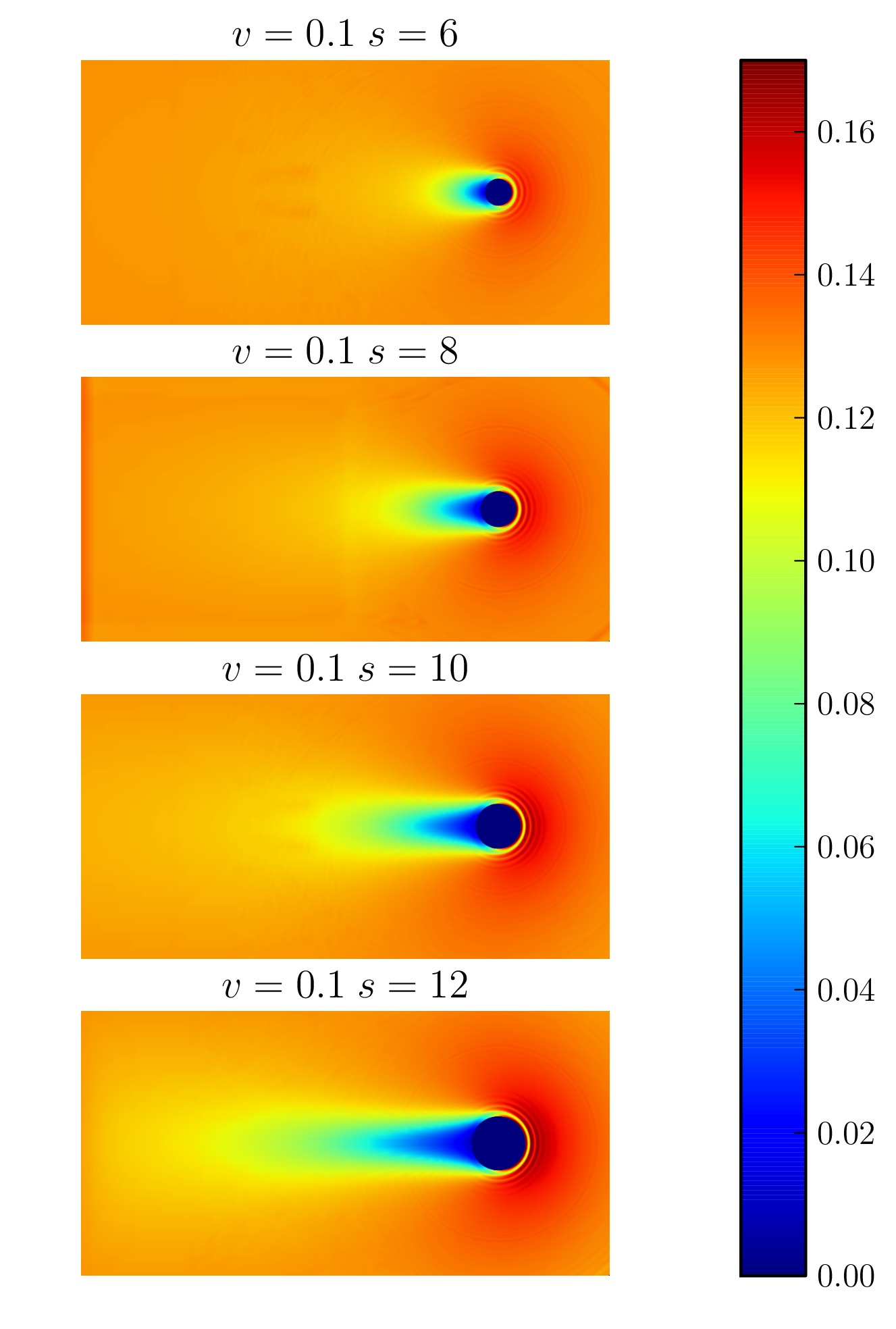}
\caption{Density distribution of the colloidal suspension driven out of equilibrium for a 
fixed dimensionless velocity $v=0.1$. 
Results are shown for relative tracer sizes (from top to bottom) 
$S=\frac{R_{c}}{R}=6, 8, 10, 12$. The volume fraction is $\varphi=0.4$ and the 
attraction parameter is $B=12.5$. 
A non-linear relation between bubble length and tracer size is apparent.}\label{sizes}
\end{figure}

The phenomenon of drying is well known for systems of attractive particles at planar hard walls; 
the density profile loses its 
oscillatory character and a layer of gas develops, which becomes infinitely thick at coexistence. 
This occurs because the particles composing the liquid 
are attracted to each other more than they are to the wall. 
Planar drying was first observed in computer simulations 
\cite{abraham,sullivan} and later using a mean field DFT similar to that employed here \cite{tarazona} (see also \cite{henderson}). 
The behaviour of the drying layer at curved substrates is complicated by the fact that the area of the interface between gas and liquid phases changes with the layer thickness. 
Due to the surface tension between gas and liquid, which acts against the increase of interfacial area, thick gas layers become energetically unfavorable (for detailed studies in this direction see 
\cite{gelfand,upton,bieker}). 
The drying layer around our circular tracer remains finite all the way up to coexistence. 


\subsection{Cavitation bubbles in bulk}
\label{Two dimensional cavity}

\begin{figure}[b!]
\includegraphics[scale=0.22]{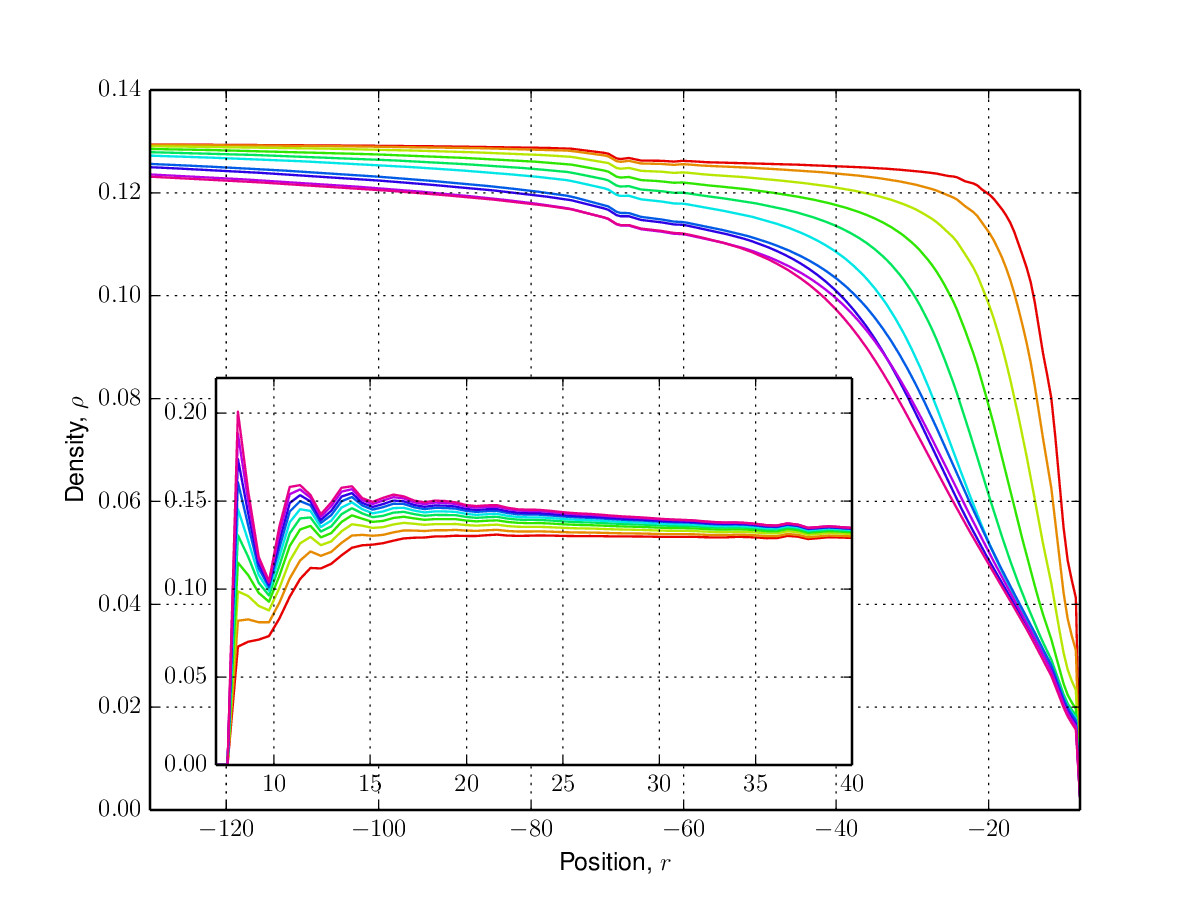}
\caption{Linear cut through the density profile behind the cavitating tracer particle 
for different dimensionless velocities. From $v=0.01$ (red) to $v=0.10$ (purple) in steps of $0.01$. 
The position is measured from the center of the tracer in units of the particles radius R. In the inset we show the region in front of the tracer (direction of the motion).
The size of the tracer is here $S=8$.}\label{otto}
\end{figure}

In Ref.\cite{micro} we found that bubble-like regions of the colloid-poor phase can form when 
a tracer is pulled through a suspension close to coexistence. 
We focus here on the influence of the tracer size upon this phenomenon. 
For a fixed dimensionless velocity $v=0.1$, we investigate the extent of the bubble 
for different values of the contact radius $R_{c}$. 
Our numerical calculations show that as the tracer size is increased the growth of the cavitation 
bubble has a nonlinear dependence on the tracer size.  
In Fig. \ref{sizes} we show steady state density distributions about tracers with relative sizes 
$S=R_c/R=6, 8, 10, 12$. 
Although the general form of the cavitation bubble remains very similar as the tracer size is increased, 
the bubble length increases significantly. 
The same observation holds for different choices of steady state velocity. 
In addition, as the tracer radius is increased the oscillatory packing at the front of the tracer 
becomes more pronounced. This is simply due to the fact that the colloidal particles take longer 
to move around a larger tracer - on the length scale of the colloids the tracer surface appears 
flatter as the tracer radius is increased. 



\begin{figure}[t!]
\includegraphics[scale=0.22]{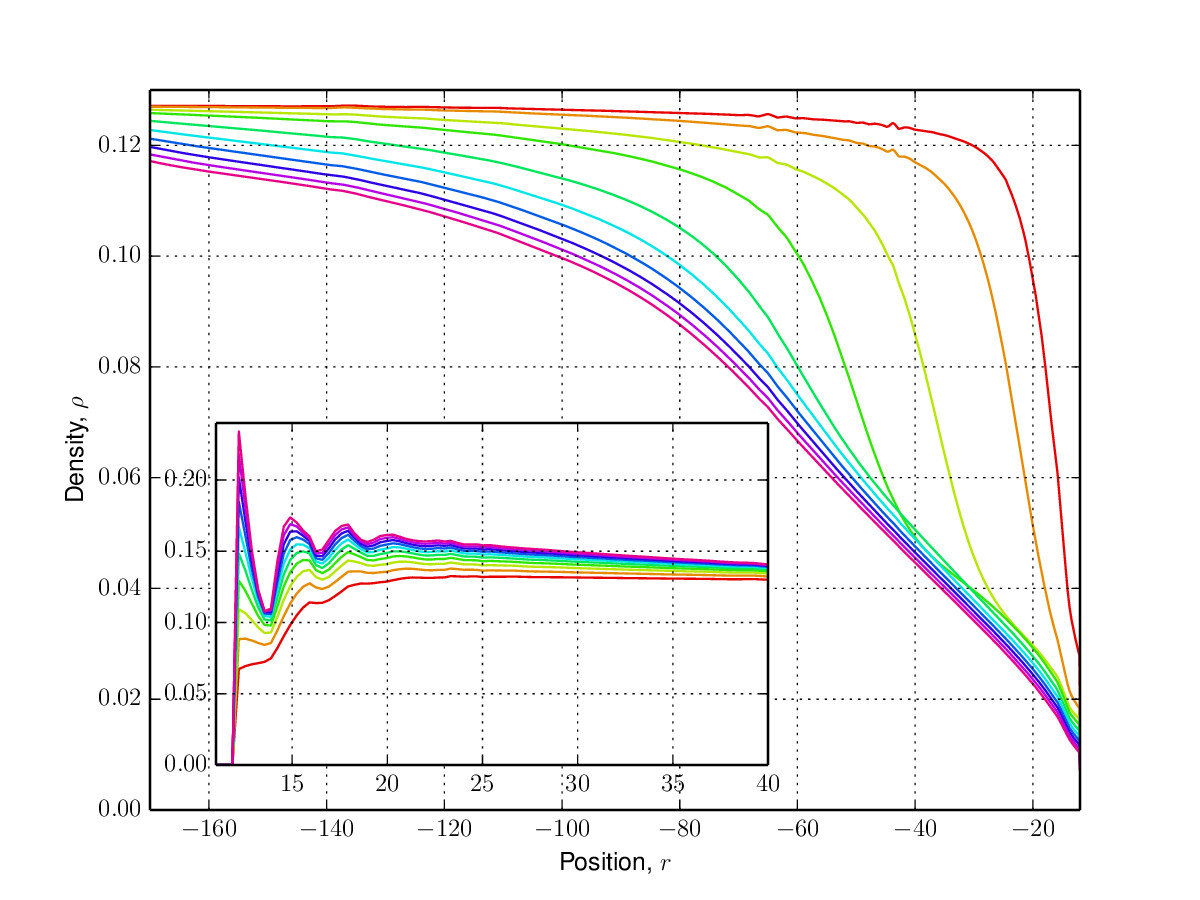}
\caption{Linear cut through the density profile behind the cavitating tracer particle 
for different dimensionless velocities. From $v=0.01$ (red) to $v=0.10$ (purple) in steps of $0.01$. 
The position is measured from the center of the tracer in units of the particles radius R. In the inset we show the region in front of the tracer (direction of the motion). The relative size of the tracer is $S=12$.}\label{dodici}
\end{figure}

In Figures \ref{otto} and \ref{dodici} we show one-dimensional slices through the two-dimensional 
density profile, following a line from the center of the tracer back through the bubble. 
In Fig.~\ref{otto} we investigate various dimensionless velocities for a tracer of fixed 
size $S=8$, whereas Fig.~\ref{dodici} shows data for the larger tracer size $S=12$.
In both cases we observe two distinct regimes as the velocity is increased. 
For example, in the case where $S=8$, in the first regime, for $0<v<0.07$, the bubble develops quite rapidly as the velocity in increased, 
indicating the sensitivity of the system at the chosen thermodynamic statepoint to small changes 
in tracer velocity. 
For  larger velocities the general form of the bubble remains rather similar, although the tail 
continues to lengthen somewhat as the velocity is increased. 
As each of these profiles is in the steady-state, they are the result of a dynamic equilibrium 
between the void formation as particles are swept away by the tracer and gradient diffusion as 
particles diffuse back into the bubble from the bulk. 
By choosing the thermodynamic statepoint to be close to the binodal we greatly facilitate 
the bubble formation by slowing the rate of colloidal diffusion from the bulk into the void 
left behind  the tracer. For the sake of completeness, we report on both figures the density in front of the tracer as an inset, where the density is piling-up and the oscillations are getting more pronounced as a function of the tracer velocity.



\begin{figure}[t!]
\includegraphics[scale=0.11]{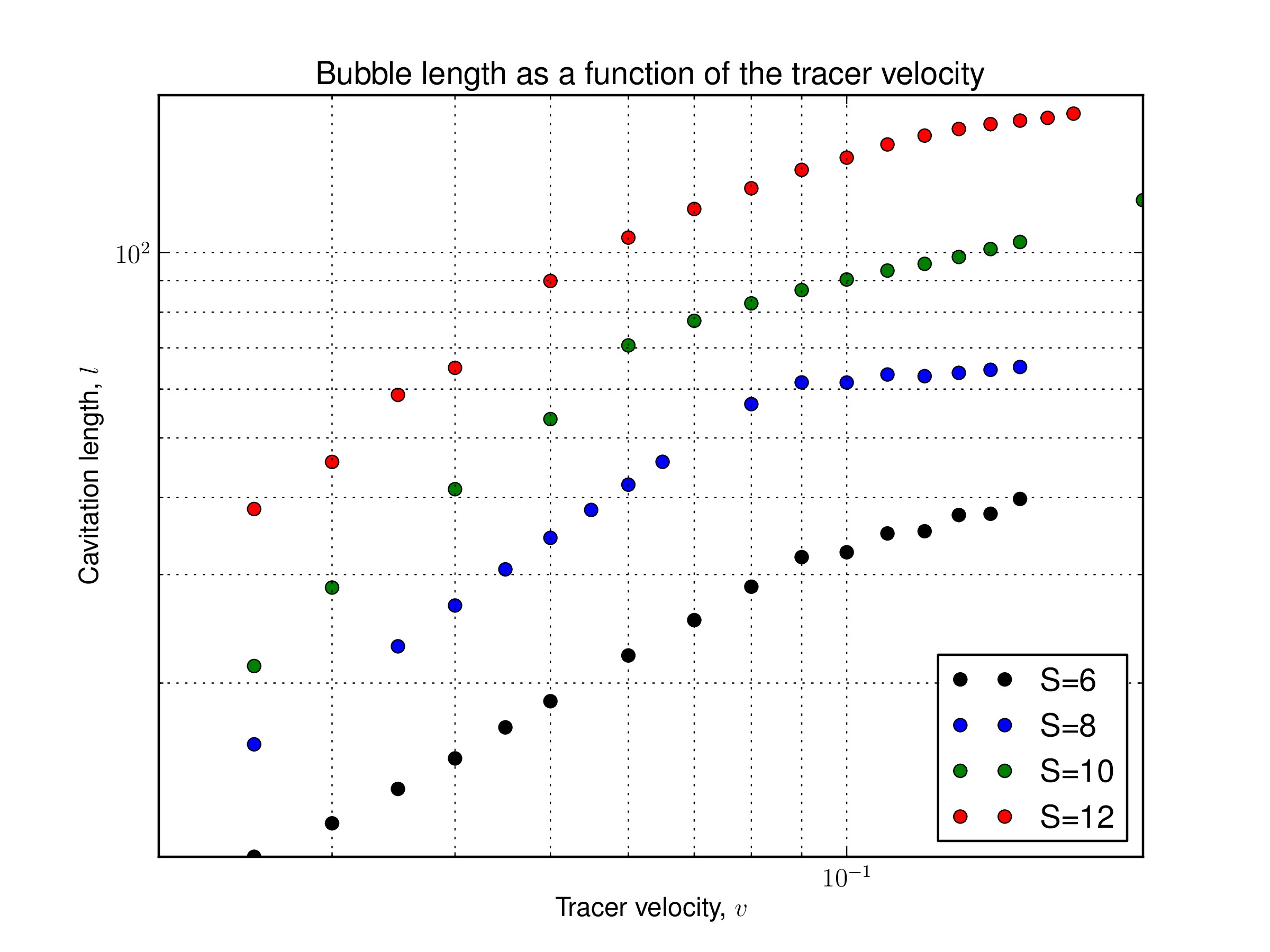}
\caption{Cavitation length as a function of the colloidal flux velocity for tracers of relative sizes $S=6, 8, 10, 12$. For each size, we can see two different regimes, which are closely linear, on the log-log scale.}\label{log-log}
\end{figure}

One can quantify the length of a cavitation bubble by defining a bubble size $l$ as the 
distance between the surface of the tracer and the point at which the colloidal 
density reaches the value of $90\%$ of the bulk density.  
In Fig.~\ref{log-log} we show this quantity as a function of dimensionless velocities for 
tracer particles with relative sizes $S=\frac{R_{c}}{R}=6,8,10,12$. 
By plotting the data on a logarithmic scale we can distinguish the two regimes of growth 
discussed above; there is a relatively clear transition from one growth exponent to another.   
The transition point between the two regimes occurs at smaller velocities for bigger test 
particle sizes, indicating that cavitation effect which develops for small velocities 
saturates more rapidly for the larger tracers.

\subsection{Cavitation close to a wall}
\label{Flat wall capillarity}

The interaction between drying (wetting) layers at two spatially separated substrates is a complex problem 
which has recieved attention in the literature at various points in time 
(see e.g.~\cite{solvent_mediated}). 
One of the 
primary motivations in studying such situations has been to gain an understanding of either 
`bridging' or `pinch off' transitions which occur when the layers of gas (liquid) phase around 
the substrates either merge together or detach from one another.

\begin{figure}[t!]
\includegraphics[scale=0.25]{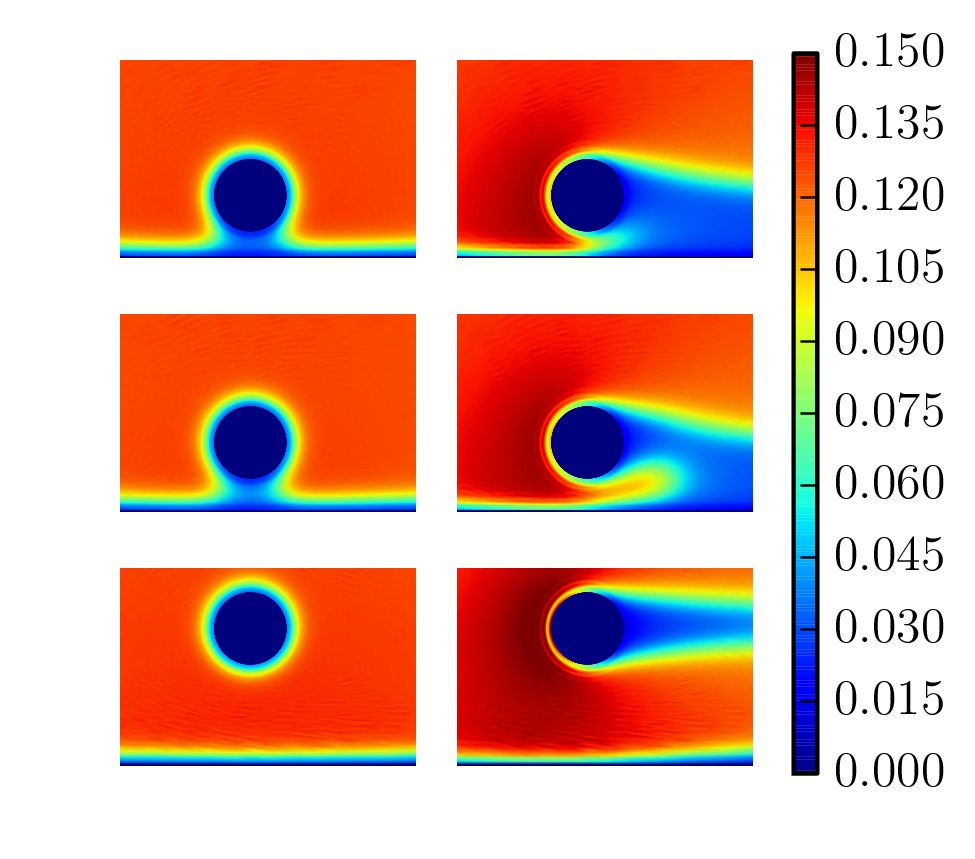}
\caption{Hard tracer particle in square-well fluid with attraction parameter $B=12.7$ with an occupied volume fraction $\varphi=0.3884$. Distances between the surface of the tracer and the flat wall are respectively, from top: 12R, 15R and 45R. On the left, the flux is zero and on the right, the flux has a velocity $v=7\times 10^{-4}$. Here the test particle radius is $R_t=15R$.}\label{wall}
\end{figure}

\begin{figure}[t!]
\includegraphics[scale=0.25]{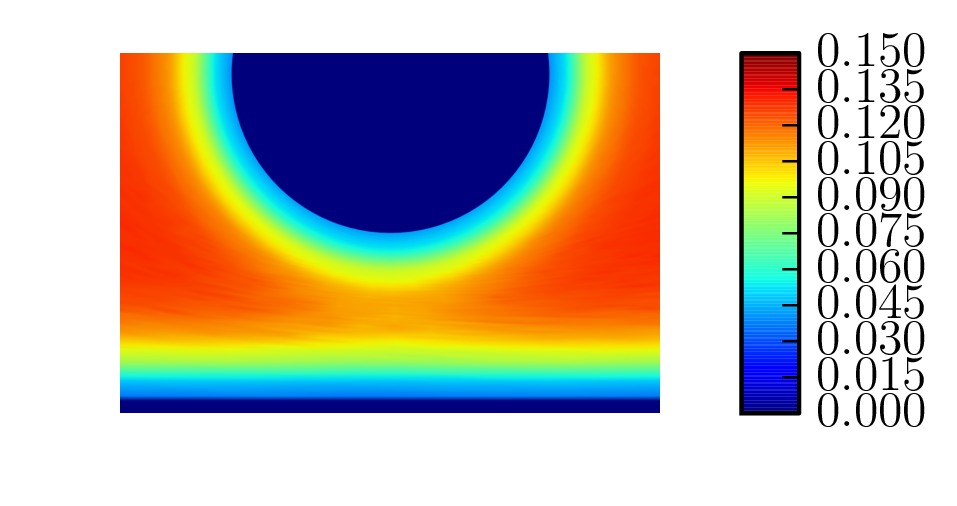}
\caption{The equilibrium density profile for a tracer-substrate separation where the bridge of low 
density colloidal phase is just visible. Note the slight distortion of the density around the tracer.}\label{ddft_zoom}
\end{figure}

The present numerical methods are rather well suited to addressing such problems, as the grid 
can be fine-tuned in critical regions between the interacting substrates.  
In the left column of Fig.~\ref{wall} we show equilibrium density profiles about a tracer of 
radius $15R$ at various distances from a planar substrate. As the separation between tracer 
and substrate is increased the bridge of colloid poor phase becomes elongated and eventually 
breaks into two distinct drying layers. 
In Fig.~\ref{ddft_zoom} we show a close-up of the density profile for a situation where the bridge has 
almost, but not quite, formed. The density around the tracer becomes slightly distorted from circular 
symmetry and some density depletion can be seen in the region between tracer and substrate.  


We now consider the analogous situation under the presence of external shear flow, given 
by Eq.~\eqref{gradient_flux}. We find that the low density bridge is very sensitive to the 
shear flow - only very small values of the dimensionless velocity are required to generate 
large deviations between the steady state and equilibrium (no shear) density. For a fixed 
flow velocity we increase the distance between tracer and wall. 
For small wall-tracer separations a very extensive bridging region is formed between the cavitation 
bubble and the drying layer at the wall. As the separation is increased the bridge starts to diminish 
and the cavitation bubble begins to detach from the wall (middle picture, right column) before, at 
still larger separations, the tracer and wall density profiles become essentially independent of 
each other. We note that very long computational times are required to obtain these steady state 
distributions and that great care must be taken to avoid numerical artifacts.

\section{Brownian Dynamics Simulations}
\label{Brownian dynamics simulations}

In order to test  in simulation the principle of colloidal cavitation predicted by 
our DDFT we implement a standard BD simulation for a system of Lennard-Jones particles 
described by a pair potential
\begin{equation}
\phi_{LJ}(r)=4\epsilon\left[\left(\frac{d}{r}\right)^{12}-\left(\frac{d}{r}\right)^6\right],
\end{equation}
where $\epsilon$ the attraction parameter and $d$ the diameter. 
The system is composed of $N=580$ particles in a square box of side-lengths 
$L_x=79R$ and $L_y=52.7R$. 
The repulsive tracer exerts a force on the colloidal bath of the form: 
$F_t(r)=24 R_C^{24}/r^{25} - 12R_C^{12}/r^{13}$. 
The tracer has a relative size of $R_t=3.5R$. The dimensionless velocity is taken as $v_f=4.2$. 
Both $k_BT$ and the diffusion coefficient, $D_0$, were set equal to unity. 
The integration time step $dt=10^{-6}$.

\begin{figure}[t!]
\includegraphics[scale=0.74]{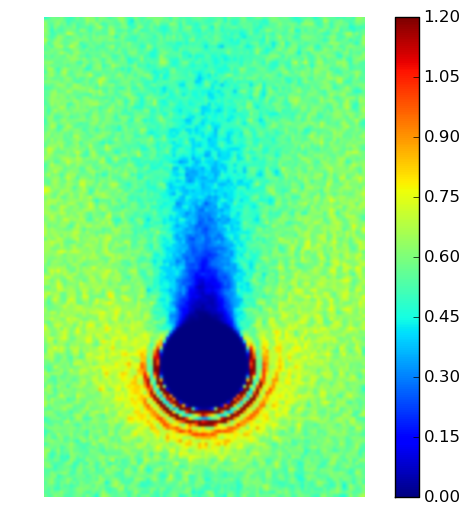}
\caption{Lennard-Jones 12-6, $\epsilon=0.1$, $\varphi=0.46$. In front of the tracer particle we can see the packing effect, due to the relative motion between the bath particles and the tracer itself. Behind it, a depleted region with a length of around $l=4R_t$ is formed, where $R_t=3.5R$.}\label{sim_tracer_1}
\end{figure}
%

\begin{figure}[t!]
\includegraphics[scale=0.74]{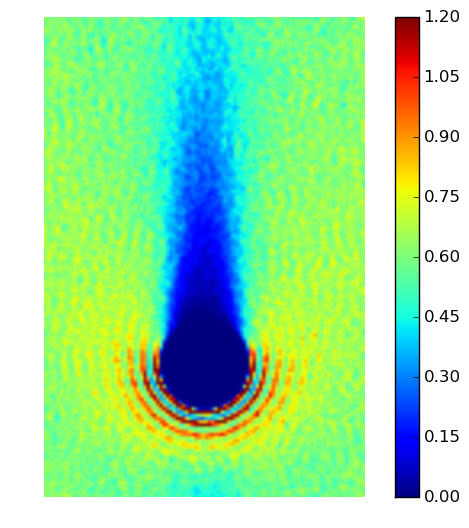}
\caption{Lennard-Jones 12-6, $\epsilon=1.4$, $\varphi=0.46$. In front of the tracer particle we can see the packing effect, due to the relative motion between the bath particles and the tracer itself. Behind it, a depleted region with a length of more then $l=6R_t$ is formed, where $R_t=3.5R$.}\label{sim_tracer_2}
\end{figure}

In Fig.~\ref{sim_tracer_1} we show the steady state density profile for a system of Lennard-Jones particles 
with an attraction parameter $\epsilon=0.1$ 
and a volume fraction of $\varphi=0.46$. At this state point the system is not phase separating. 
By increasing the attraction parameter up to $\epsilon=1.4k_BT$ (Fig.~\ref{sim_tracer_2}), which is close to 
the phase boundary, we can see an increase in the length of the cavitating region. 
The phenomenology we observe here is consistent with that observed in our DDFT calculations.

\begin{figure}[t!]
\includegraphics[scale=0.55]{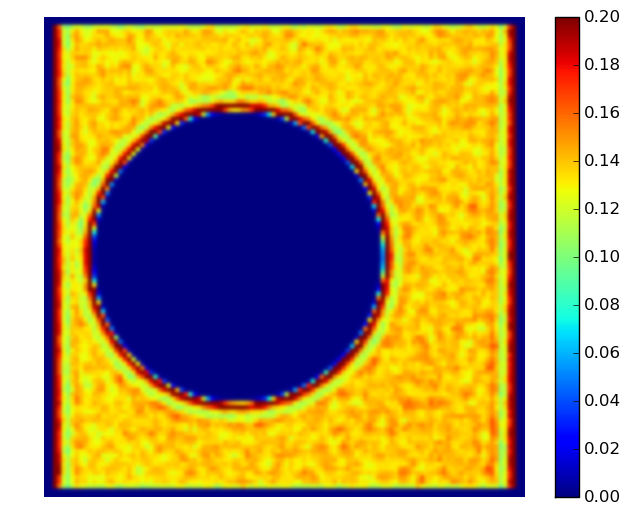}
\caption{The number of particles $N=250$, the occupied volume fraction $\varphi=0.46$, the parameter $\epsilon=0.1$, the cutoff $r_{cut}=1.12d$, the tracer radius $R_t=15R$, distance between substrate left and tracer surface $D_L=5R$ and the distance to the right wall $D_R=15R$.}\label{sim_HS}
\end{figure}

\begin{figure}[t!]
\includegraphics[scale=0.55]{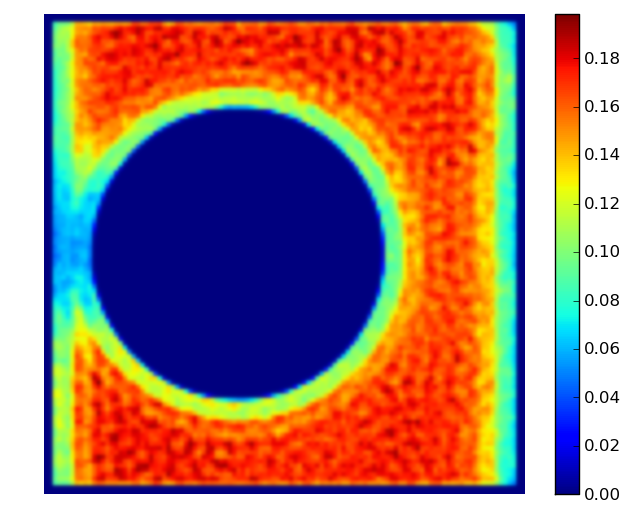}
\caption{The number of particles $N=250$, the occupied volume fraction $\varphi=0.46$, the attraction parameter $\epsilon=2$, the cutoff $r_{cut}=2.5d$, the tracer radius $R_t=15R$, distance between substrate left and tracer surface $D_L=5R$ and the distance to the right wall $D_R=15R$.}\label{sim_LJ}
\end{figure}

In section \ref{Flat wall capillarity} we studied a hard tracer particle in an attractive square well bath, 
close to a repulsive substrate and observed that a low density bridge can form between the wall and the 
tracer.  
As a phenomenological test of the bridging phenomenon we now consider a repulsive tracer close to a repulsive 
substrate in two different colloidal systems: 
(i) a soft repulsive system of particles interacting via a cut and shifted Lennard-Jones potential 
(chosen for technical reasons),
(ii) an attractive system interacting via a standard (12-6) Lennard-Jones potential.   
In Fig.~\ref{sim_HS} we show the density for the soft repulsive system and we observe packing effects 
close to the hard substrate, as well as around the hard tracer.
In Fig.~\ref{sim_LJ} we calculate the density profile at the same bulk density, but now using the attractive 
Lennard-Jones system at a state point close to phase separation. 
We clearly observe the development of a bridge between the two drying layers. 

\section{Discussion}\label{discussion}

We have considered two related problems. The first concerns a purely repulsive tracer particle 
moving with constant velocity, immersed in a bulk square-well suspension. 
For a state point in the liquid phase, close to the binodal, a region of depleted density appears 
behind the tracer. The length of this bubble does not grow linearly as a function of the tracer 
diameter. For a fixed tracer size, we then studied the length of the cavitation bubble 
as a function of the 
tracer velocity. We identified two regimes in the bubble formation. In the first regime the length of the 
bubble increases strongly as a function of the relative velocity, then, in the second one, the increment is much more contained. 
Using Brownian dynamics we have shown that 
the flow induced cavitation is a phase transition induced phenomenon which can be observed in 
simulation. This provides the first qualitative verification of our DDFT predictions. 
Using the same model system we then addressed the situation when the tracer is close to a substrate 
and subject to a shear flow. 
Both the tracer and the wall develop drying layers which interact with each other and with the external 
flow field in a complicated way. 

By increasing the distance between tracer and substrate the capillarity became thinner and longer and eventually broke into two distinct drying layers. Considering the analogous situation under the presence of external shear flow, we found that the low density bridge is very sensitive to external deformations. We fixed the flow velocity and increased the distance between tracer and substrate. Initially, an extensive bridging between the cavitation bubble and the drying layer at the wall was formed. By increasing the distance this bridge started to diminish and eventually the cavitation bubble detached completely from the wall, forming two independent colloid poor regions.

We finally proposed a phenomenological test using Brownian Dynamics simulations. We first demonstrated that the motion of the tracer can induce a local liquid-vapour phase transition. In addition to that we showed results proving that, for a system composed by a repulsive tracer and a repulsive wall in a thermodynamic state point such that both are surrounded by a gas region, a bridge between the two drying layers is formed, when the distance between the two bodies is small enough.

\subsection*{Acknowledgements}
We thank the Swiss National Science Foundation for financial
support under the grant number 200021-153657/2.


\begin{thebibliography}{}


\bibitem{mason}
T.G.~Mason, D.A.~Weitz, Physical Review Letters. {\bf 74} 1250 (1995). 

\bibitem{macosko}
C.W.~Macosko, {\it Rheology: Principles, Measurements, and Applications} 
(Wiley, 1994).

\bibitem{larson}
R.G.~Larson, {\it The Structure and Rheology of Complex
Fluids} (Oxford University Press, New York, 1999).

\bibitem{mewis}
J.~Mewis and N.J.~Wagner, {\it Colloidal suspension rheology} 
(Cambridge University Press, 2012).

\bibitem{gazuz}
I.~Gazuz, A.M.~Puertas, Th. Voigtmann and M.~Fuchs,
Phys. Rev. Lett. {\bf 102} 248302 (2009).

\bibitem{gnann}
M.V.~Gnann, I.~Gazuz, A.M.~Puertas, M.~Fuchs and Th. Voigtmann, 
Soft Matter {\bf 7} 1390 (2011).

\bibitem{micro}
J.~Reinhardt, A.~Scacchi, J.M.~Brader,
J.Chem.Phys. {\bf 140} 144901 (2014).



\bibitem{marconi_tarazona_1}
U.~M.~B.~Marconi and P.~Tarazona, J. Chem. Phys. {\bf 110} 8032 (1999).

\bibitem{marconi_tarazona_2}
U.~M.~B.~Marconi and P.~Tarazona, J. Phys.: Condens. Matter {\bf 12} A413 (2000).

\bibitem{archer_evans}
A.J.~Archer and R.~Evans, J. Chem. Phys. {\bf 121} 4246 (2004).

\bibitem{archer_rauscher}
A.J. Archer and M. Rauscher, J. Phys. A: Math. Gen. {\bf 37} 9325 (2004).


\bibitem{hopkins}
P. Hopkins, A.J. Archer and R. Evans, J. Chem. Phys. {\bf 131} 124704 (2009).

\bibitem{malijevski}
A. Malijevsky, Mol. Phys. {\bf 113} 1170 (2015).

\bibitem{malijevski_parry}
A. Malijevsky and A.O. Parry Phys. Rev. E {\bf 92} 022407 (2015).






\bibitem{dhont_book}
J.~K.~G. Dhont, {\it An introduction to dynamics of colloids} (Elsevier, Amsterdam, 1996).


\bibitem{evans79}
R.~Evans, Adv. Phys. {\bf 28},  143  (1979). 







\bibitem{kb1}
J.M. Brader and M.Kr\"uger,
Mol.Phys. {\bf 109} 1029 (2011).

\bibitem{kb2}
M.Kr\"uger and J.M. Brader,
EPL {\bf 96} 68006 (2011)

\bibitem{skb}
A.~Scacchi, M.~Kr\"uger 
and J.M.~Brader,                                                                                
J.Phys.:Condens.Matter {\bf 28} 244023 (2016).

\bibitem{oettel}
R.~Roth, K.~Mecke and M.~Oettel, J.Chem.Phys. {\bf 136} 181101 (2012).

\bibitem{archer_chacko}
A.J. Archer, B. Chacko and R. Evans, J. Chem. Phys. {\bf 147} 034501 (2017).

\bibitem{roland_review}
R.~Roth, J-Phys.:Condens.Matter {\bf 22} 063102 (2010).

\bibitem{Bangerth2007}
W.~Bangerth, R.~Hartmann and G.~Kanschat,
ACM Trans. Math. Softw. {\bf 33} 24 (2007).

\bibitem{Geuzaine2009}
C.~Geuzaine and J.-F.~Remacle, 
International Journal for Numerical Methods in Engineering, 
{\bf 79} 1309 (2009).





   









\bibitem{abraham}
F.F.~Abraham, J.Chem.Phys. {\bf 68} 3713 (1978).

\bibitem{sullivan}
D.E.~Sullivan, D.~Levesque and J.J.~Weiss, J.Chem.Phys. {\bf 72} 1170 (1980).

\bibitem{tarazona}
P.~Tarazona and R.~Evans,
Mol.Phys. {\bf 52} 847 (1984). 

\bibitem{henderson}
J.R.~Henderson and F.~van Swol, 
Mol.Phys. {\bf 56} 1313 (1985). 

\bibitem{gelfand}
M.P.~Gelfand and R.~Lipowsky, Phys.Rev.B {\bf 36} 8725 (1987).

\bibitem{upton}
P.J.~Upton, J.O.~Indekeu and J.M.~Yeomans, Phys.Rev.B, {\bf 40} 666 (1989).

\bibitem{bieker}
T.~Bieker and S.~Dietrich, Physica A, {\bf 252} 85 (1998). 

\bibitem{solvent_mediated}
A.J.~Archer, R.~Evans, R.~Roth, M.~Oettel
J. Chem. Phys. {\bf 122}, 084513 (2005).









\end{thebibliography}
\end{document}